\documentclass{article}
\usepackage{emulateapj}
\usepackage{apjfonts}
\usepackage{timesfonts}
\usepackage{psfig}

\begin{document}

\def\cf{{\it cf.\/}}
\def\ie{{\it i.e.\/}}
\def\eg{{\it e.g.\/}}
\def\etal{{\it et al.\/}}
\def\theta{\vartheta}
\def\phi{\varphi}

\title{Illuminated, and enlightened, by GRB 991216}

\author{M. Vietri}
\affil{Universit\`a di Roma 3, Via della Vasca Navale 84, I--00147 Roma, 
Italy;e--mail: vietri@fis.uniroma3.it}

\author{G.Ghisellini and D. Lazzati\altaffilmark{1}}
\affil{Osservatorio Astronomico di Brera, Via E. Bianchi 46 I--23807 
Merate, Italy; e--mail: gabriele@merate.mi.astro.it}

\author{F. Fiore and L. Stella}
\affil{Osservatorio Astronomico di Roma, Via Frascati 33,I--00040 Monteporzio 
Catone, Italy; e--mail: fiore,stella@coma.mporzio.astro.it}

\altaffiltext{1}{Present address: Institute of Astronomy, Madingley Road,
Cambridge CB3 0HA, UK; e--mail: lazzati@ast.cam.ac.uk}

\begin{abstract}
We consider models for the generation of the emission line recently 
discovered in the X--ray afterglow spectrum of several bursts, and especially
of GRB 991216 observed by $Chandra$.
These observations suggest the presence of 0.1--1 solar masses of 
iron in the vicinity of the bursts.
We show that there are strong geometrical and kinematical constraints on
the line emitting material.
We discuss several classes of models, favoring one where the line photons 
are produced by reflection off the walls of a wide funnel of semi--opening 
angle $\theta \approx 45^\circ$ excavated in a young (a few months old) 
supernova remnant made of $\approx 10 M_\odot$ with radius 
$6\times 10^{15}$ cm, and $\approx 1 M_\odot$ of iron, 
providing strong support for the SupraNova model. 

\end{abstract}


\section{Introduction}

There are now five bursts with evidence for large amounts of material 
around the site of the explosion: four 
(GRB 970508, Piro et al., 1999;
GRB 970828, Yoshida et al., 1999; 
GRB 991216, Piro et al., 2000, hereafter P2000; 
GRB 000214, Antonelli et al., 2000) 
show an emission feature during the afterglow, and one 
(GRB 990705, Amati et al., 2000) 
displays an edge in absorption during the burst itself. 
The observations of GRB 991216 by $Chandra$, begun 37~hr after the burst 
and lasting for about 3 hr, show a 3.49$\pm$0.06 keV line 
significant at a  $\sim 4\sigma$ level. 
If the line is identified with the recombination K$\alpha$ line 
from H--like iron (6.97 keV), the inferred redshift is $z= 1.00\pm 0.02$
consistent with the largest optical absorption redshift system ($z = 1.02$) 
(Vreeswijk et al., 2000). The line, resolved by $Chandra$, displays a width 
corresponding to $0.05c$ (P2000). 
The line and the 2--10 keV continuum fluxes were  
$\sim 1.6\times 10^{-13}$ and $\sim 2.4\times 10^{-12}$ erg
s$^{-1}$ cm$^{-2}$, respectively, implying (for $z=1.02$) an emission rate of 
iron line photons of $\dot{N}_{\rm line} \simeq 4\times 10^{52}$ s$^{-1}$, 
luminosity $L_{\rm line} \simeq 4\times 10^{44}$ erg s$^{-1}$
and total energy $E_{\rm line}\sim 3 \times 10^{49}$ erg
(assuming steady emission for $40/(1+z)$~hr, 
with $H_0=75$ km s$^{-1}$ Mpc$^{-1}$ and $q_0=0.5$).
If each iron atom produces $k$ line photons, the required iron
mass is $M_{\rm Fe} \simeq 195\ k^{-1}$~ M$_\odot$. 
Bringing this mass down to $\sim 0.1$~M$_\odot$ implies $k>2000$ which in turn
implies a limit on the recombination time
{\footnote{ From the formulae given in Verner \& Ferland (1996),
the recombination time can be approximated by
$t_{\rm rec}\sim 12.7 T_7^{3/4}n_{10}^{-1}$ s 
in the temperature range $10^6<T<10^8$K.}}.
This translates into the condition 
$n_{\rm e} > 10^{10} T_7^{3/4}$ cm$^{-3}$; the electron
temperature $T=10^7 T_7$ K comes from interpreting the broad bump, 
marginally detected by $Chandra$, at 4.4$\pm$0.5 keV as 
the recombination continuum of H--like iron (9.28 keV). 
Analogously to GRB~950708 (Lazzati et al. 1999), the line detection 
implies the presence of $\approx 0.5 M_\odot$ of pure
iron in the vicinity of the burster. 

Observation of the line $t_{\rm obs}$ after the burst implies that the 
line emitting material must be within $\sim ct_{\rm obs}/(1+z)$
from the burst site: this region must then be {\it compact}, 
contain $\sim 0.1 M_\odot$ only in iron, but nevertheless be 
optically thin to electron scattering, such that Comptonization
does not broaden the line beyond the observed width. 
We call this the {\it size problem}.
 
Furthermore, if we interpret the line width as due to the velocity
of a supernova remnant, the limit on the size allows to estimate the age
of the remnant: for GRB 991216,  $\sim 15$~days.
At this time (Fig. 1), cobalt nuclei outnumber nickel and iron 
nuclei and the line would be produced mainly by cobalt, not iron, at an 
energy $\epsilon = 7.5/(1+z)$ keV.
We call this the {\it kinematic problem}. 

We discuss in section 2 the constraints implied by the line observation; 
we then present in section 3 three scenarios which could produce the observed 
line.
Two of them are found to be viable only by adopting ad hoc assumptions or 
through fine tuning, while a third scenario, 
the ``wide funnel" scenario, appears to be more promising.
In section 4 we discuss our results.

\section{Physical and geometrical conditions}

The detection of emission features in the afterglow spectra of GRBs some 
hours after the GRB event poses a strong constraint on the location of the 
line emitting material.
If the line is detected after $t_{\rm obs}$ from the burst explosion
the material must be located within a distance $R$ given by
\begin{equation}
R\, \le\, {ct_{\rm obs} \over 1+z} \, {1\over 1-\cos\theta}\, \simeq \,
{ 1.1\times 10^{15}\over 1+z}  \, {t_{\rm obs} \over10\, {\rm h} } \,\, 
{1\over 1-\cos\theta}\,\, {\rm cm},
\label{eq:rag}
\end{equation}
where $\theta$ is the angle between the line emitting material and the 
line of sight at the GRB site. This limit has important implications, 
because the scattering optical depth is large,
\begin{equation}
\tau_{\rm T}\, = \, {\sigma_{\rm T} M \over 4\pi R^2 \mu m_{\rm p}}
\ge \, 54 \, { (M/M_\odot)(1+z)^2(1-\cos\theta)^2 \over \mu \, 
(t_{\rm obs}/10\, {\rm h})^2} \ \ ,
\end{equation}
(where $\mu=1$, 1.2 and 2 for pure hydrogen, solar composition, 
and no hydrogen, respectively), and, for a remnant radial velocity 
of $v=10^9v_9$ cm s$^{-1}$
the time elapsed from the SN is $t_{\rm SN} \simeq 
12.5 (t_{\rm obs}/10{\rm hr}) / [(1+z)(1-\cos\theta)v_9]$ days, 
implying that most $^{56}$Co nuclei (and a fraction of $^{56}$Ni)
have not yet decayed 
to $^{56}$Fe (half--lives of 77.3 and 6.08 days, respectively).
In Fig. 1 we show normalized abundances of nickel, cobalt and
iron as a function of time from the SN explosion.

To produce the line by frequent recombinations and ionizations,
a sufficiently high ionizing flux and a large  number of H--like iron ions, 
FeXXVI, are required. Efficient use of these ions requires that 
the photoionization optical depth for FeXXVI, 
$\tau_{\rm FeXXVI} \ga 1$.
Using  photoionization cross section 
$\sigma_{\rm Fe XXVI}\sim 1.2\times 10^{-20}$ cm$^{-2}$,
and abundance ratio $\xi= M_{\rm FeXXVI}/ M_{\rm Fe}$, we have
\begin{equation}
\tau_{\rm FeXXVI}\, = {\sigma_{\rm Fe XXVI} \xi M_{\rm Fe} \over
56 m_{\rm p} 4\pi R^2}\, =\,
2\times 10^3 {\xi \over R^2_{15}} \, {M_{\rm Fe} \over 0.1\, M_\odot} \ .
\end{equation}
A strict {\it lower} limit on the ionizing continuum power
comes from the line flux itself.
For GRB 991216, for which there is no {\it rebursting} in the X--ray
afterglow (contrary to GRB 970508), 
assuming an illumination time of  20 hr,
we require $E_{\rm ion}>3\times 10^{49}$ erg.
Even in the case of an illuminating continuum not directly visible by us, 
we require that the scattered flux produced by the line
emitting material does not affect the afterglow emission, i.e.
it must contribute less than 10$^{-12}$ erg cm$^{-2}$ s$^{-1}$ to the
observed flux, giving the {\it upper} limit 
$\min(1,\tau_{\rm T}) E_{\rm ion}<1.8\times 10^{50}$ erg.

\subsection{Compton broadening and Thomson optical depth}

If the emitting atoms are diluted in an electron cloud, 
the fraction of unscattered flux $ = [1-\exp(-\tau_{\rm T})]/\tau_{\rm T}$,
while if the scattering cloud surrounds a central source
the fraction $= \exp(-\tau_{\rm T})$.
If the scattering electrons have a temperature in the $\sim 10^7$~K 
range (according to the indication that the recombination  
continuum observed in GRB991216 may be 
broad, P2000), the average energy shift of 
line photons scattered only once is 
$\Delta E/E \approx 0.063 (kT/1\, {\rm keV})^{1/2}$
(Pozdnyakov, Sobol \& Sunyaev 1983), 
{\it already broader than the observed line width}. 
Multiple scatterings would further broaden the line, making 
it increasingly difficult to detect against the X--ray continuum. 
If the line of GRB 991216 were made by photons 
escaping unscattered from a medium with  $\tau_{\rm T}>1$, the 
problem with the line flux would be exacerbated, a factor 
$\tau_{\rm T}$ more iron being required. 

An alternative possibility, that the electron temperature $\ll
10^7$~K, say $\sim 10^6$ K expected if the ionizing 
continuum were steep and extended to low energies, might
conjure with a suitable value of $\tau_{\rm T}$ to yield, through
Compton broadening, the observed line width.
In this case, however, the centroid of the line would be 
redshifted (line photons are mostly backscattered by 
colder electrons) and the recombination continuum narrower.
Due to poor statistics, the recombination continuum of 
GRB 991216 cannot place a firm constraint on the electron temperature,
but the models in Section 3 are largely independent 
of the presence and width of the recombination emission continuum. 

\section{The models}

We can roughly divide the models into {\it transmission} and
{\it reflection} models.
If we see the line in reflection, line photons come from the 
layer with $\tau_{\rm FeXXVI}=$~several\footnote{What we call FeXXVI 
might well be CoXXVII or NiXXVIII, see section 2.}. 
For an iron abundance greater than the solar value, 
in this layer $\tau_{\rm T}\le 1$.
In general, transmission models require $\tau_{\rm T}< 1$
and $\tau_{\rm FeXXVI}\sim 1$, while reflection models require 
$\tau_{\rm FeXXVI}>\tau_{\rm T}$.
Each iron atom must produce $\sim 2000$ iron photons,
and this requires an electron density larger than $10^{10}$ cm$^{-3}$.

The large equivalent widths of the X--ray line in
GRB 991216 ($EW=0.5$ keV) and, especially, GRB 970508
($EW\sim 1$ keV), GRB 970828 ($EW\sim 3$ keV) and GRB 000214
($EW\sim 2$ keV) favor models in which the line originates in 
reflection, not transmission.
In reflection, in fact, the ionizing flux is always efficiently
reprocessed in line photons in the $\tau_{\rm Fe}\sim1$ layer, while
in transmission this happens only for a particular tuning of the FeXXVI and
the free electron densities.

Three alternative geometries (Fig. 2) producing lines in reflection 
are described below. 

In reflection models, we can derive the photon line luminosity
by estimating the volume $V_{\rm em}$ effectively contributing to 
the line emission, and assuming a given iron mass.
If the layer contributing to the emission has $\tau_{\rm T}\sim 1$
(in order to avoid excessive Compton broadening of 
the line), and in this layer $\tau_{\rm FeXXVI}\sim$a
few (to efficiently absorb the continuum),
we have $V_{\rm em}=S/(\sigma_{\rm T} n_{\rm e})$, where $S$ is the 
emitting surface. 
The line emission rate from $V_{\rm em}$ is then
\begin{equation}
\label{rate}
\dot N_{\rm Fe} = {N_{\rm Fe}\over t_{\rm rec}} =
{Sn_{\rm Fe} \over 1.3\times 10^{11} T_7^{3/4} \sigma_{\rm T} } \sim 
3\times 10^{53} {(M_{\rm Fe}/M_\odot) \over T_7^{3/4} \Delta R_{15}}
\quad {\rm s^{-1}} \ .
\label{eq:ndot}
\end{equation}
where we assumed that the {\it total} volume is $V=S \Delta R$
(slab or shell geometry). 
Eq.~\ref{eq:ndot} shows that the total iron mass must be a sizable 
fraction of a solar mass to obtain the observed line flux
($4\times 10^{52}$~s$^{-1}$), but less than the largest observed values, 
$0.9 M_\odot$ for SN 1998bw (Type Ic, Sollerman et al., 2000) and 
SN1991T, (Type Ia, Filippenko et al., 1992). 
Eq. \ref{rate} all by itself establishes 
that the line emitting material must be a SNR: no other known astrophysical
object contains so much iron, not even the largest known stars (though these
fail by a factor of a few only).

Eq. \ref{eq:ndot} has another important implication:
it fixes $M/R$, and since $R$
is fixed (for a given geometry, see below) 
by the time--delay, the cloud density
is also fixed, to within an order of magnitude. 
Solutions with densities much exceeding (or
much below) $\approx 10^{10}$ cm$^{-3}$ are excluded.

\subsection{The wide funnel}

If the GRB explodes within the young remnant of a supernova,
we must consider two possibilities: a plerionic and a shell remnant.
In the first case, we consider a wide funnel excavated 
in a young plerionic remnant.
This can solve the {\it size problem}, since  it extends to large radii 
but, at the same time (Eq.~\ref{eq:rag}), can maintain the time--delay 
contained because it is naturally built close to the polar axis 
(assumed close to the line of sight). 
The geometry is sketched in Fig.~2a.
Fixing the line photons rate (Eq. \ref{rate}) yields $R = 6\times 10^{15}$ cm, 
and thus an opening angle $\theta = 48^\circ$, to fit the time--delay.
For any reasonable SN composition, these parameters imply 
$n_{\rm e} > 10^{10}$~cm$^{-3}$. 
The funnel walls are probably not straight (like in a cone geometry),
but curved instead, like the surface of the coffee in a
cup when it is spun up, and can be 
illuminated by the ionizing flux from a central source.
Assuming a cone geometry for simplicity, we rewrite 
Equation~\ref{eq:ndot} as
\begin{equation}
\dot N_{\rm Fe} \, =\, 3.3\times 10^{52} 
{(M_{\rm Fe}/M_\odot) \over T_7^{3/4}  (R_{15}/6)} \tan{\theta}
\quad {\rm s^{-1}} \ . 
\label{eq:ndot2}
\end{equation}
where in this case $R$ (the radius of the filled remnant) 
has been used instead of $\Delta R$ and $\tan{\theta}$
takes into account the geometry of the funnel.
This is a lower limit, since a parabolic geometry has a larger funnel 
surface and we neglected the (likely) density stratification inside 
the remnant. 

Since the funnel's normal is {\it not} parallel to the incident photon's 
momentum, radiation pressure exerts a force parallel to the 
surface on the layer $\tau_{\rm T} \leq 1$ (which reflects 
isotropically photons arriving from a specific direction). 
Calling $\phi$ the angle between the funnel's normal and the 
incident photon's arrival direction the absorbed fluence 
$E_{\rm ion}$ accelerates the funnel layer to a speed 
$v_{\rm f} = \sin\phi(2 E_{\rm ion}/M_{\rm layer})^{1/2} \simeq  
10^4 E_{\rm ion, 50}^{1/2}\sin\phi$~km~s$^{-1}$ when
$R=6\times 10^{15}$~cm. Thus, ablation by radiation pressure propels 
the reflecting layer to velocities like  those seen in GRB991216 by P2000.

The funnel model can thus explain the observed line broadening 
even in a relatively old SNR. For a remnant with 
$v_{\rm ej} = 3000$~km~s$^{-1}$, a typical value for type II SNe, we
find a remnant age $t_{\rm SNR} \sim 230$~days, enough
to turn most $^{56}$Ni and $^{56}$Co into $^{56}$Fe (Fig.~1). 
The emitting iron is then boosted to higher velocity by 
the ionizing flux, circumventing the kinematical problem.

\subsection{Back of the remnant}

We consider now a shell remnant, characterized by a wide opening, so that 
a substantial fraction of its inner face is visible (Fig.~2b),
and illuminated by burst and afterglow and, more likely, by the shock 
due to the fireball impact (B\"ottcher 2000).
Line photons are produced in the layer $\tau_{FeXXVI}\sim$~few. 
As before, the remnant may have been 
expanding slowly, until accelerated by burst photons.
Since we see the remnant's internal part, the {\it size} limit 
discussed in Section 2 is tighter, and we detect mainly 
redshifted photons.
For a remnant distance $R=10^{15}$ cm and velocity 5,000
km s$^{-1}$, the age is $\sim 20$~days, 
close to the $^{56}$Co peak (Fig. 1).
However, the 7.5 keV $^{56}$Co line would be observed 
with a substantial redshift (and therefore mimic 
a $^{56}$Fe 6.97 keV line), if the burst photons increased 
the expanding velocity to $\sim 20,000$~km~s$^{-1}$
(implying that the burst
emitted a few times $10^{52}$~erg sideways, for a 10 $M_\odot$ remnant).
In this model the observed line width of 15,000 km s$^{-1}$
requires that only a limited range of projections
of the radial expansion velocity vectors are observed.
This solution is not convincing since it involves fine tuning
between the energy produced sideways and the mass of the remnant,
in order to have the correct expansion velocity, even though it 
explains naturally the different centroid energy of the line observed in 
GRB~970508: since the line is redshifted by shell expansion, we expect 
a distribution of observed energies as a consequence of different 
velocities of the remnant.

\subsection{Back--illuminated equatorial material}

We now explore the possibility of a simultaneous GRB--``supernova"
explosion.
Consider a scenario in which the GRB ejects and accelerates 
a small amount of matter in a collimated cone, while a large amount of 
matter is instead ejected, at sub--relativistic speeds,
along the progenitor's equator.
For illustration, assume equal amounts of energy, 
$10^{52}$ erg, 
in both directions, and 1~M$_{\odot}$ expelled along the equator.
Then $v_{\rm eq}/c =0.148 E_{52}^{1/2} (M/M_\odot)^{-1/2}$.
Massive star progenitors are always surrounded
by dense material from strong winds with mass loss rates
$\dot m_{\rm w}= 10^{-5} \dot m_{\rm w, -5}$
and velocity $v_{\rm w}=10^7v_{\rm w,7}$.
This wind scatters back a fraction of the photons produced by the bursts
and its afterglow (Thompson \& Madau 2000),
,nd, in return it is also accelerated until it reaches relativistic
velocities, when scattering efficiency decreases.
Thus assume that each electron (and the associated proton) 
scatters photons only until it reaches $\Gamma=2$, i.e. until it has 
scattered a total energy of $m_{\rm p}c^2$ (i.e. $\sim$2,000
photons of 0.5 MeV each).
In this case the scattered luminosity
$L_{\rm scatt}$ is constant, since there are an equal number 
of electrons in a shell of constant width $\Delta R$
(for a density profile $\propto R^{-2}$).
This luminosity is of the order:
\begin{equation} 
L_{\rm scatt} \, \sim \, m_{\rm p} c^2\,  
{\dot m_{\rm w}\over m_{\rm p} v_{\rm w}/c}\, =\, 
1.8\times 10^{45}\,
{\dot m_{\rm w, -5}\over v_{\rm w,7}} \,\,\,\, {\rm erg\, s^{-1}} \ .
\end{equation}
Scattered photons illuminate the expanding 
equatorial matter after a time $2R/c$, producing X--ray line emission
mostly due to K$\alpha$ emission from H--like $^{56}$Ni at 8.1 keV.
Even if the large expanding velocity makes the transverse Doppler
shift important, this requires a fine tuned velocity to redshift
the 8.1 keV $^{56}$Ni line into 6.97 keV,
for which reason we consider this scenario unlikely.

\section{Conclusions}

X--ray line emission features from GRBs' afterglows give 
us information on the nature of their progenitors by imposing
constraints on models, the most severe being how to arrange a
large amount of line emitting material around the GRB site, but without
large Thomson scattering opacities.
A further limit comes from the line width of $Chandra$'s observation 
in GRB 991216. These observations require a large iron mass (Eq. 
\ref{rate}), such as is seen only in SNe. 

We have presented three distinct models:
two (the ``back of the remnant" and 
``back illuminated equatorial material" models) 
require that the observed emission line is produced  
by redshifted cobalt or nickel (instead of iron), and need 
fine tuning and/or ad hoc assumptions.
The ``wide funnel" model, instead, solves the size problem, and the 
acceleration of the line emitting material by grazing incident 
photons solves the kinematic problem, allowing time for 
cobalt to decay to iron.
It implies that the GRB progenitors are massive stars 
exploded as supernovae months before the burst, contaminating 
the circum--burst environment with iron rich material.
This two--step process, and the time--delay between the two steps, are 
predicted by the SupraNova scenario (Vietri \& Stella, 1998).
Alternative explanations invoke extremely iron enriched
massive winds (Weth et al. 2000).

Future observations 
may address issues such as the presence of cobalt and nickel
lines in the spectrum, the age of the remnant, 
the line profile, the geometry and kinematics of the 
emitting region, the presence of more than one line,
the ionization state, the time evolution of the line and
edges, all indicative of the characteristics of the illuminator.
Together, this information may determine whether the line emitting material
originates in a normal, albeit rare, supernova remnant or if a more exotic
explosive phenomenon is required.

\acknowledgements
We thank Frits Paerels and Martin J. Rees for useful discussions.

\bigskip
\bigskip
\bigskip

\psfig{file=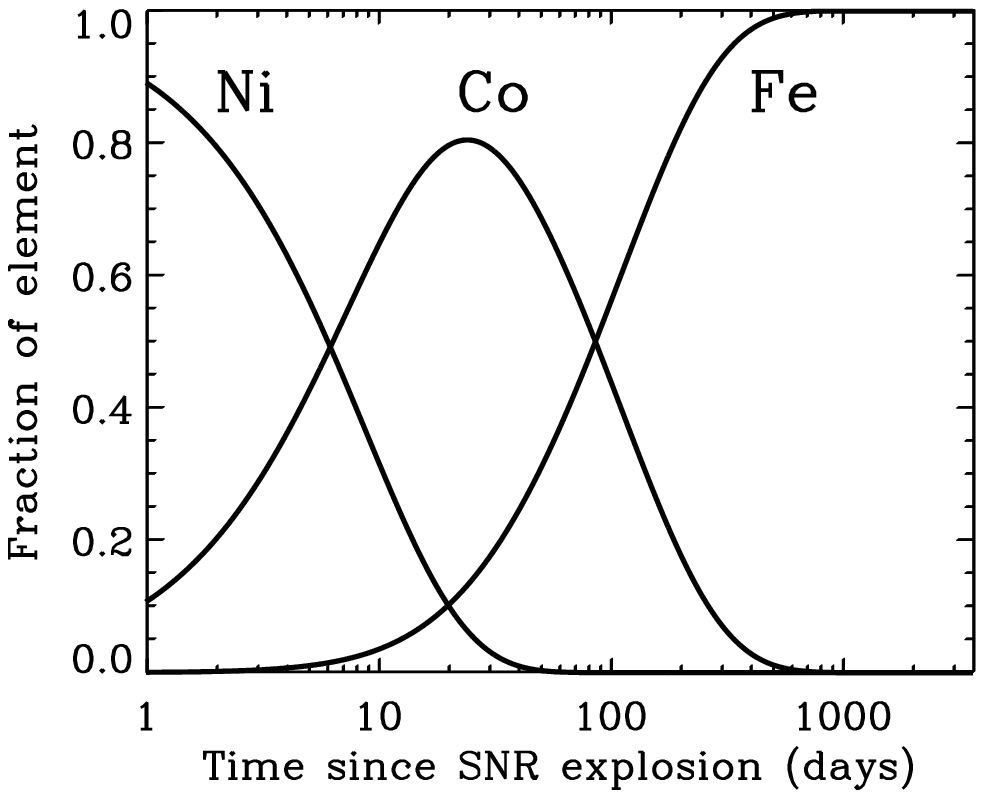}
\figcaption[vietri_fig1.eps]{The 
normalized abundances of $^{56}$Ni, $^{56}$Co and $^{56}$Fe as a 
function of time from the creation of $^{56}$Ni.
The neutral, He--like, H--like and recombination edge energies are (in keV):
$^{56}$Ni: 7.478, 7.806, 8.102, 10.775;
$^{56}$Co: 6.930, 7.242, 7.526, 10.012;
$^{56}$Fe: 6.404, 6.701, 6.973, 9.278.}

\bigskip
\bigskip
\bigskip

\psfig{file=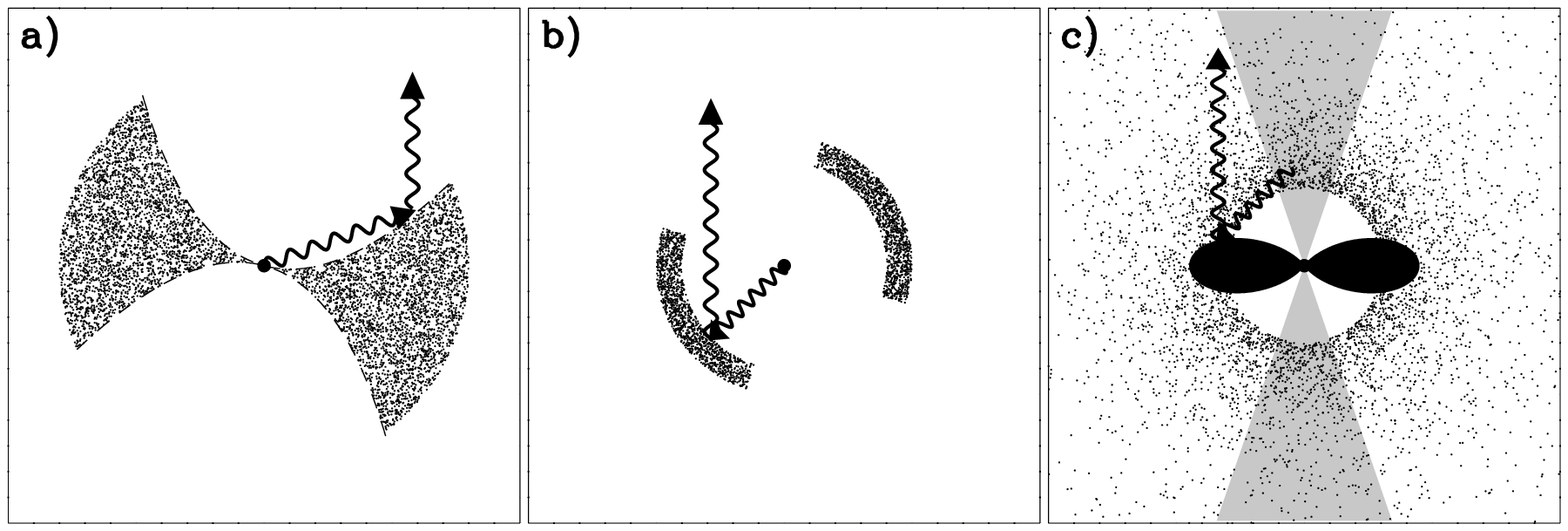}
\figcaption[vietri_fig2.eps]
{Cartoon of the three models discussed in this paper.
In panel a) a funnel excavated in a supernova remnant produces 
the iron line in reflection, by material accelerated to $\sim$10,000
km s$^{-1}$ by the strong radiation pressure.
In panel b) we see the internal parts of a young and receding
supernova remnant, in panel c) some equatorial material exploding 
simultaneously with the burst is illuminated by burst and afterglow 
photons scattered by the pre--burst stellar wind.}

\end{document}